\documentclass[%
groupedaddress,
preprint,
amsmath,amssymb,
aps,
]{revtex4-2}

\usepackage{graphicx}
\usepackage{dcolumn}
\usepackage{bm}

\usepackage[figuresright]{rotating}
\usepackage{multirow}
\usepackage{array}
\usepackage{threeparttable}
\usepackage{indentfirst}%
\usepackage{ulem}
\usepackage{subcaption}
\usepackage{float}

\draft

\begin{document}





	\title{$\alpha$-decay half-lives and $\alpha$-cluster preformation factors of nuclei around $N=Z$ line}
	\author{Jing Li$^{1,2}$}
	\author{Shan He$^{1,2}$}
	\author{Yueqing Li$^{1,2}$}
	\author{Weiwei Wang$^{1,2}$ }
	\author{Yanzhao Wang$^{1,2,3,4}$ }
 	 \email{yanzhaowang09@126.com}
	\author{Jianzhong Gu$^{4}$ }
	 \email{gujianzhong2000@aliyun.com}
	\affiliation{$^1$ Department of Mathematics and Physics, Shijiazhuang Tiedao University, Shijiazhuang 050043, China\\
		$^2$ Institute of Applied Physics, Shijiazhuang Tiedao University, Shijiazhuang 050043, China\\
		$^3$ Hebei Key Laboratory of Physics and Energy Technology, North China Electric Power University, Baoding 071000, China\\
		$^4$ China Institute of Atomic Energy, P. O. Box 275 (10), Beijing 102413, China}

\date{\today}

\begin{abstract}
\noindent In this work, a microscopic effective nucleon-nucleon interaction based on the Dirac-Brueckner-Hartree-Fock $G$ matrix starting from a bare nucleon-nucleon interaction is used to explore the $\alpha$-decay half-lives of the nuclei near $N=Z$ line.  Specifically, the $\alpha$-nucleus potential is constructed by doubly folding the effective nucleon-nucleon interaction with respect to the density distributions of both the $\alpha$-cluster and daughter nucleus. Moreover, the $\alpha$-cluster preformation factor is extracted by a cluster formation model. It is shown that the calculated half-lives can reproduce the experimental data well. Then, the $\alpha$-decay half-lives that are experimentally unavailable for the nuclei around $N=Z$ are predicted, which are helpful for searching for the new candidates of $\alpha$-decay in future experiments. In addition, by analyzing the proton-neutron correlation energy and two protons-two neutrons correlation energy of $Z=52$ and $Z=54$ isotopes, the $\alpha$-cluster preformation factor evolution with $N$ is explained. Furthermore, it is found that the two protons-two neutrons interaction plays more important role in $\alpha$-cluster preformation than the proton-neutron interaction. Meanwhile, proton-neutron interaction results in the odd-even effect of the $\alpha$-cluster preformation factor.
\end{abstract}


\maketitle
\newpage
\section{Introduction}\label{sec1}
$\alpha$-decay has been recognized as a powerful tool for extracting the structure and properties of unstable nuclei. Experiments show that a large number of $\alpha$-radioactivity occur in heavy nuclei~\cite{Audi2003}. Recent years, the mass region dominated by $\alpha$-decay has been continuously extended with the rapid development of experimental techniques. Therefore, a number of superheavy $\alpha$ emitters have been successfully synthesized by cold-, warm-, and hot-fusion-reactions~\cite{Ogan99,Mori09,Hof09}. Besides, the $\alpha$-radioactivity was detected from some medium-mass nuclei approaching the $N=Z$ line, including $^{105-110}$Te~\cite{Sch79,Liddi06,Sew06,Sew02,Janas05}, $^{108-113}$I~\cite{Moe91,Kirch77}, $^{109-112}$Xe~\cite{Sch79,Liddi06,Schar81,Roeckl78}, $^{112, 114}$Cs~\cite{Roeckl78,Cart12} and $^{114}$Ba~\cite{Mazz02}.

Recently, a new $\alpha$-decay chain $^{108}$Xe$\to ^{104}$Te$\to ^{100}$Sn was observed experimentally by the Argonne National Laboratory, including the measurements of the $\alpha$-particle kinetic energy and $\alpha$-decay half-lives of new $\alpha$ emitters $^{108}$Xe [$E_{\alpha }$ = 4.4(2) MeV, $T_{1/2}=58_{-23}^{+106}\mu$s] and $^{104}$Te [$Q_{\alpha }$ = 4.9(2) MeV, $T_{1/2}<18$ ns] being produced by the fusion-evapotation reaction $^{54}$Fe($^{58}$Ni, $4n$)$^{108}$Xe~\cite{Aura18}. This is not only the first time to observe $\alpha$-radioactivity to a heavy self-conjugate nucleus $^{100}$Sn but also helpful to reveal the physical mechanism of $\alpha$-clustering near the nuclei of $N=Z$ line. Therefore, studies on the $\alpha$-decay properties around $N=Z$ line are attracting more and more attention~\cite{Wan2021,Wang14,Wang14b,Ghar22,Kane03,Xu2006,Xu2019,Clark2020,Deng2022,Bai2018,Poenaru06}. Because these $\alpha$ emitters are important for studying the structure and properties of nuclei approaching the $N=Z$ line as well as the shell effect around the $N=Z$=50~\cite{Wan2021,Wang14,Wang14b,Ghar22,Kane03,Xu2006,Xu2019,Clark2020,Deng2022,Bai2018,Poenaru06}.



It is well known that $\alpha$-decay is described as a quantum tunneling model firstly proposed by Gamow~\cite{Gamow28} and by Condon and Gumey~\cite{Condon28} in 1928. On the basis of this picture, various macroscopic and microscopic approaches have been developed~\cite{Buck93,Lovas98,Mous01,Li2010,Wang2025,Qi2024,Qi2025,Delion2016,LXH2016,LXH2017,LXH2025,Qian2023,Wang2015,pei2006,pei2007,Ghodsi2024,Chowdhury07,Bhattacharya08,Mohr2017,Ismail}. Among these approaches, the $\alpha$-cluster potential in the nuclear surface plays an important role in determining the half-life. Generally, the $\alpha$-cluster potential is constructed by double folding the density distributions of $\alpha$ and daughter nuclei with an effective nucleon-nucleon ($n-n$) interaction. In fact, there exists many versions of the effective $n-n$ interactions on the basis of different physical origins, such as the Skyrme interaction~\cite{pei2006,pei2007,Ghodsi2024,Ward2013}, the Michigan-3-Yukawa interaction~\cite{Chowdhury07,Bhattacharya08,Chow08} and the meson exchange $n-n$ interaction~\cite{Mohr2017,Ismail,Bhuyan2018}. However, there remains a degree of uncertainty in the current understanding of the $n-n$ interaction. Therefore, when calculating the half-life of $\alpha$-decay using different versions of effective $n-n$ interactions, certain discrepancies in computational accuracy arise~\cite{pei2006,pei2007,Ghodsi2024,Chowdhury07,Bhattacharya08,Mohr2017,Ismail}. Consequently, to reproduce the experimental $\alpha$-decay half-lives well, employing a reasonable effective $n-n$ interaction is crucial. Conversely, the $\alpha$-decay experimental data also provides an excellent ground for testing $n-n$ effective interactions.

In Ref.~\cite{Schiller01}, an effective microscopic $n-n$ interaction through effective meson exchange was obtained within the Dirac-Brueckner-Hartree-Fock (DBHF) $G$ matrix starting from a bare $n-n$ interaction. This type of $n-n$ interaction has been successfully used to investigate the nuclear structure and scattering~\cite{Ma2002,Rong2006,Ma2008,Zou2008}. Next, the effective $n-n$ interaction within the DBHF model was used to estimate the $\alpha$-decay half-lives of superheavy nuclei. It was found that the experimental $\alpha$-decay half-lives were reproduced well by the approach~\cite{Zhang2010}. Because a series of $\alpha$ emitters approaching the $N=Z$ line were discovered, it is interesting to extend the DBHF approach to study the $\alpha$-decay of this mass region to further test the effective $n-n$ interaction. Besides, previous studies indicated that there was a larger $\alpha$-particle formation probability around the $N=Z$ nuclei owing to a strong overlap between the neutron and proton Fermi energies~\cite{Wan2021,Wang14,Wang14b,Ghar22,Kane03,Xu2006,Xu2019,Clark2020,Deng2022,Bai2018,Poenaru06}. However, the evolution of $\alpha$-cluster preformation probability with $N$ has not been fully understood. Recent years, the cluster formation model (CFM) is usually used to extract the $\alpha$-cluster preformation probability~\cite{Wan2021,Ahmed2013,Ahmed2015}. Therefore, in this work we will use DBHF approach and CFM to study the $\alpha$-decay properties of nuclei around the $N=Z$ line and the evolution of $\alpha$-cluster preformation probability with $N$.

This article is organized as follows. Section II presents the theoretical frameworks of the DBHF and CFM. The numerical results and discussions are given in Section III. The last section presents a conclusion.

\section{Theoretical Framework}

The total interaction potential $V(R)$ between the $\alpha$-cluster and the daughter nucleus is given by

\begin{equation}  \label{VR}
	V\left ( R \right ) =V_{N}\left ( R \right )  +V_{c}\left ( R \right )+V_{l}\left ( R \right ).
\end{equation}
Here, $V(R)$ includes the nuclear potential $V_{N}\left( R \right )  $, the Coulomb potential $V_{c}\left( R \right )  $, and the centrifugal potential $V_{l}\left( R \right )  $, respectively. The quantity $R$ is the distance between the mass center of the $\alpha$-cluster and that of the daughter nucleus.

For the nuclear potential $V_{N}\left( R \right )  $, it is from the effective $n-n$ interaction doubly folded with nucleon density distributions of both the $\alpha$-particle and the daughter nucleus~\cite{Zhang2010}

\begin{equation}  \label{VN(R)}
	V_{N}\left ( R \right ) =\lambda \iint \rho _{\alpha } \left ( \mathbf{r_{1}}  \right ) \rho _{d}\left ( \mathbf{r_{2}}   \right )V_{\mathrm{eff} }^{\mathrm{ILDA} }\left ( \varepsilon , \rho , \mathbf{s}  \right )\mathrm{d}\mathbf{r_{1}} \mathrm{d}\mathbf{r_{2}},
\end{equation}
where \textit{V}$_{\text{eff}}^{\text{ILDA}}$ is the effective $n-n$ interaction by considering the improved local density approximation (ILDA) obtained within the DBHF $G$ matrix starting from a bare $n-n$ interaction~\cite{Schiller01,Zhang2010,Zou2008}, which is written as

\begin{equation}  \label{Veff}
	V_{\mathrm{eff} }^{\mathrm{ILDA} } \left ( \varepsilon , \rho , \mathbf{s}  \right ) =g\left ( \mathbf{s}  \right )V_{\mathrm{eff} } ,
\end{equation}

\noindent where
\begin{equation}  \label{g(s)}
	g\left ( \mathbf{s}  \right ) =\left ( t_{R}\sqrt{\pi }   \right ) ^{-3} \exp \left ( -\frac{\mathbf{s^{2}}  }{t_{R}^{2}  }  \right ) ,
\end{equation}

\noindent and $\mathbf{s} =\mathbf{R} +\mathbf{r_{2}}  -\mathbf{r_{1}}  $ corresponds to the relative vector between interacting nucleon pairs, where $\mathbf{r_{1}}$ and $\mathbf{r_{1}}$ are the coordinates of the nucleon in the center-of-mass frame of the $\alpha$-particle and the daughter nucleus, respectively, and $t_{R} $ is a range parameter. In Eq. \eqref{VN(R)}, $\lambda$, $\rho$ and $\varepsilon$ are the renormalization factor, local nuclear matter density and kinetic energy in the nucleon-nucleon center-of-mass frame, respectively. For $V_{\mathrm{eff} } $, it is the effective $n-n$ interaction without the ILDA and is calculated by the following Thomas-Fermi approximation: $V_{\mathrm{eff} } =U_{\mathrm{eff} } /\rho$. Here, \textit{U}$_{\text{eff}}$ is the nucleon self-energy in the nuclear medium.\\\vspace{-3ex}

In the DBHF approach, the real part of the nucleon self-energies is parametrized in polynomial expansions, which is expressed as
\begin{equation}  \label{ReUeff}
	\text{Re}U_{\mathrm{eff} } \left ( \rho , E \right ) =\sum_{i=1}^{3}\sum_{j=1}^{3}a_{ij}\rho ^{i}\varepsilon ^{j-1},
\end{equation}
where $E$ refers to the total energy of the nucleon in the nuclear medium, that is $E=\varepsilon +M$. Here, $M$ is the rest mass of nucleons. The coefficients $a_{ij} $ are listed in Table \ref{tab1}~\cite{Zhang2010,Zou2008}.\\\vspace{-3ex}

\begin{table}[!h]
	\centering
	\caption{Values of the coefficients $a_{ij}$ in Eq. \eqref{ReUeff}.}
	\label{tab1}
	\begin{tabular}{cccc}
		\hline\hline
		$a_{ij}$ & 1      & 2 & 3  \\
		\hline
		1   & $-0.797\times 10^{3}$  & $0.222\times 10^{1}$  & $-0.825\times 10^{-3}$   \\
		2   & $0.366\times 10^{4}$       & $-0.438\times 10^{1}$  & $-0.696\times 10^{-2}$   \\
		3   & $-0.515\times 10^{4}$      & $0.584\times 10^{1}$   & $0.221\times 10^{-1}$    \\
		\hline\hline
	\end{tabular}
\end{table}

In  Eq. \eqref{VN(R)}, the local density $\rho$ is simply calculated by the geometric average of the density distributions for the two interacting nuclei~\cite{Zhang2010,Zou2008,Cara96}
\begin{equation}  \label{rho}
	\rho =\sqrt{\rho _{\alpha  }\left ( \mathbf{r_{1}}   \right )\rho _{\alpha } \left (\mathbf{r_{2}}  \right )  } .
\end{equation}

The nucleon density distribution of $\alpha$-particle $\rho _{\alpha } \left ( r \right ) $ is given by a Gaussian form~\cite{Satch79}
\begin{equation}  \label{rho-alpha}
	\rho _{\alpha } \left ( r \right ) =\frac{4}{\pi ^{3/2}b_{\alpha }^{3}  }\exp \left ( -\frac{r^{2} }{b_{\alpha }^{2} }  \right ),
\end{equation}
\noindent where $b_{\alpha }$=1.1932 fm.

The nucleon density distribution of daughter nucleus $\rho _{d} \left ( r \right )$ is taken by the spherically symmetric Fermi function~\cite{Zhang2010,Zou2008}
\begin{equation}  \label{rho-d}
	\rho_{d}  \left ( r \right ) =\frac{\rho _{0} }{1+\exp \left [ \left ( r-c \right ) /a \right ]  } ,
\end{equation}
\noindent where the central density $\rho _{0} =\frac{3A}{4\pi c^{3}\left ( 1+\pi ^{2}a^{2}/c^{2}    \right )  } $, the half-density radius $c=1.07A_{d}^{1/3} $, and the diffuseness $a$= 0.54 fm.

The renormalization factor $\lambda$ is obtained by the Bohr-Sommerfeld quantization condition
\begin{equation}  \label{Bohr}
	\int_{R_{1} }^{R_{2} } \sqrt{\frac{2\mu }{\hbar^{2} } \left | Q_{\alpha }-V\left ( r \right )  \right | } ~\mathrm{d}r =\left ( 2n+1 \right ) \frac{\pi }{2}  =\left ( G-L+1 \right ) \frac{\pi }{2} ,
\end{equation}

\noindent where $n$ is the node number of the radial wave function, and the global quantum number $G$=16 $\left (50< Z,~N\le  82  \right ) $~\cite{Zhang2010,Mohr00} is adopted in this work. Moreover, $R_{1} =0$ is the first turning point. The second turning point $R_{2} $ and the third turning point $R_{3} $ are determined by solving the equation $V(R)=Q_{\alpha }$.\\\vspace{-3ex}

The Coulomb potential $V_{c}\left( R \right )  $ in Eq. \eqref{VR} is obtained by
\begin{equation}  \label{Vc(R)}
	V_{c}\left ( R \right )=\begin{cases}  & \frac{Z_{\alpha }Z_{d} e^{2}  }{R} ~~~~~~~~~~~~~~~~~~~~~~~~ R> R_{c}  \\  & \frac{Z_{\alpha }Z_{d} e^{2}  }{2R_{c}  }\left [ 3-\left (\frac{R}{R_{c} }   \right )^{2}   \right ] ~~ R\le R_{c}\end{cases},
\end{equation}

\noindent where $Z_{\alpha } $ and $Z_{d }$  denote the proton number of the $\alpha$-particle and the daughter nucleus, respectively. The parameter $R_{c} =1.2A_{d }^{1/3} $ is the Coulomb radius of the daughter nucleus and $A_{d }$ is the mass number of the daughter nucleus.\\

The centrifugal potential $V_{l}\left( R \right )  $ in Eq. \eqref{VR} can be calculated by
\begin{equation}  \label{Vl(R)}
	V_{l}\left ( R \right ) =\frac{\hbar ^{2} }{2\mu }  \frac{l(l+1)}{R^{2} } ,
\end{equation}
\noindent where $l$ is the angular momentum carried by $\alpha$-cluster. $\mu$ is the reduced mass, which is denoted as $\mu=M_{\alpha }M_{d}/M_{p}$, where $M_{\alpha }$, $M_{d}$ and $M_{p}$ are the masses of the $\alpha$-particle, the daughter nucleus, and the parent nucleus, respectively.\\\vspace{-3ex}

The $\alpha$-decay half-life is calculated by
\begin{equation}  \label{half life}
	T_{1/2 } =\frac{\hbar\ln_{}{2}}{\Gamma_{\alpha}},
\end{equation}

\noindent where $\Gamma_{\alpha}$ is the $\alpha$-decay width. It can be expressed as
\begin{equation}  \label{decay width}
	{\Gamma_{\alpha}}=P_{\alpha}F\frac{\hbar^{2} }{4\mu }\exp\left [ -2\int_{R_{2} }^{R_{3} }k\left ( R \right )\mathrm{d}R    \right ].
\end{equation}

\noindent In Eq. \eqref{decay width} , $F$ is the normalization factor, which is calculated by
\begin{equation}  \label{F}
	F\int_{R_{1} }^{R_{2} } \frac{\mathrm{d}R }{2k\left ( R \right ) } =1.
\end{equation}

\noindent Here, the wave number $k(R)$ is obtained by
\begin{equation}  \label{k(R)}
	k\left ( R \right ) =\sqrt{\frac{2\mu }{\hbar ^{2 } }\left | Q_{\alpha }-V\left ( R \right )  \right |  } ,
\end{equation}
where $Q_{\alpha }$ is the $\alpha$-decay energy.


Based on the CFM approach, the preformation factor $P_{\alpha }$ of an $\alpha$-cluster in a nucleus is given by~\cite{Wan2021,Ahmed2013,Ahmed2015}
\begin{equation}  \label{preformation factor}
	P_{\alpha }=\frac{E_{f\alpha } }{E_{f\alpha }+E_{r}  } =\frac{E_{f\alpha } }{E_{tot}} ,
\end{equation}
\noindent where $E_{f\alpha }$ is the $\alpha$-cluster-formation energy associated with the interaction among the four nucleons in the $\alpha$-cluster, $E_{r}$ is the energy for the relative motion derived from $\alpha$-particle orbiting the daughter nucleus in the ground state of the parent nucleus, and $E_{tot}=E_{f\alpha }+E_{r}$ is the total energy for the $\alpha$-clustering state~\cite{Wan2021,Ahmed2013,Ahmed2015}. Generally, to calculate $P_{\alpha }$, the values of $E_{f\alpha }$ and $E_{tot}$ in Eq. \eqref{preformation factor} should be calculated by solving the corresponding Schr\"odinger equation. However, by analyzing the surface nucleon-nucleon interaction, the values of $E_{f\alpha }$ and $E_{tot}$ can both be approximately calculated from the experimental separation energies\cite{Wan2021,Ahmed2013,Ahmed2015}
\begin{equation}  \label{Efa}
	E_{f\alpha } =\begin{cases}  &2S_{p}+2S_{n}-S_{\alpha } \text{   (even-even)  } \\  &2S_{p}+S_{2n}-S_{\alpha } ~\text{ (even-odd) }  \\  &S_{2p}+2S_{n}-S_{\alpha } ~\text{ (odd-even) }  \\  &S_{2p}+S_{2n}-S_{\alpha } ~\text{ (odd-odd) } \end{cases},
\end{equation}

\begin{equation}  \label{E}
	E_{tot}=S_{\alpha }  \left ( A, Z \right ),
\end{equation}

\noindent where $S_{p}$, $S_{n}$, $S_{2p}$, $S_{2n}$, and $S_{\alpha }$ represent one-proton, one-neutron, two-proton, two-neutron, and $\alpha$-cluster separation energies, respectively. These values can be obtained by

\begin{equation}  \label{Sp}
	S_{p}\left ( A, Z \right )  =B\left ( A, Z \right )-B\left ( A-1, Z-1 \right )  ,
\end{equation}

\begin{equation}  \label{Sn}
	S_{n}\left ( A, Z \right )  =B\left ( A, Z \right )-B\left ( A-1, Z \right )   ,
\end{equation}

\begin{equation}  \label{S2p}
	S_{2p}\left ( A, Z \right )  =B\left ( A, Z \right )-B\left ( A-2, Z-2 \right )    ,
\end{equation}

\begin{equation}  \label{S2n}
	S_{2n}\left ( A, Z \right )  =B\left ( A, Z \right )-B\left ( A-2, Z \right )    ,
\end{equation}

\begin{equation}  \label{Sa}
	S_{\alpha }\left ( A, Z \right )  =B\left ( A, Z \right )-B\left ( A-4, Z-2 \right )   ,
\end{equation}
\noindent where $B(A, Z)$ is the binding energy of a nucleus with mass number $A$ and proton number $Z$. Generally, the experimental $B(A, Z)$ values are used, whose values are obtained from the AME2020 atomic mass table~\cite{AME2021a}. For the nuclei without the experimental data, the $B(A, Z)$ values are taken from the Weizs\"acker-Skyrme-4 (WS4) mass table~\cite{WS4}.\\\vspace{-3ex}

\section{Results and Discussions}

Firstly, the parameter $t_{R} $ in Eq. \eqref{g(s)} needs to be determined by the following $\chi ^{2} $ least-square fit
\begin{equation}  \label{chi}
	\chi ^{2} =\sum_{i=1}^{n}\left (\frac{T_{1/2,i }^{\mathrm{expt.}}-T_{1/2,i }^{\mathrm{cal.}} }{T_{1/2,i }^{\mathrm{expt.}}}   \right )  ^{2}  ,
\end{equation}
\noindent where $T_{1/2,i }^{\mathrm{expt.} } $ and $T_{1/2,i }^{\mathrm{cal.}} $ are the experimental and calculated $\alpha$-decay half-lives by inputting the experimental $Q_{\alpha } $ values, respectively. Within Eq. \eqref{chi}, it is found that the value of $t_{R} $ is 0.9 fm when $\chi ^{2} $ reaches a minimum, which is the optimal value in the framework of the DBHF model of this article.\\\vspace{-3ex}

\begin{figure}[!h]
	\begin{center}
		\includegraphics[width=10.0cm]{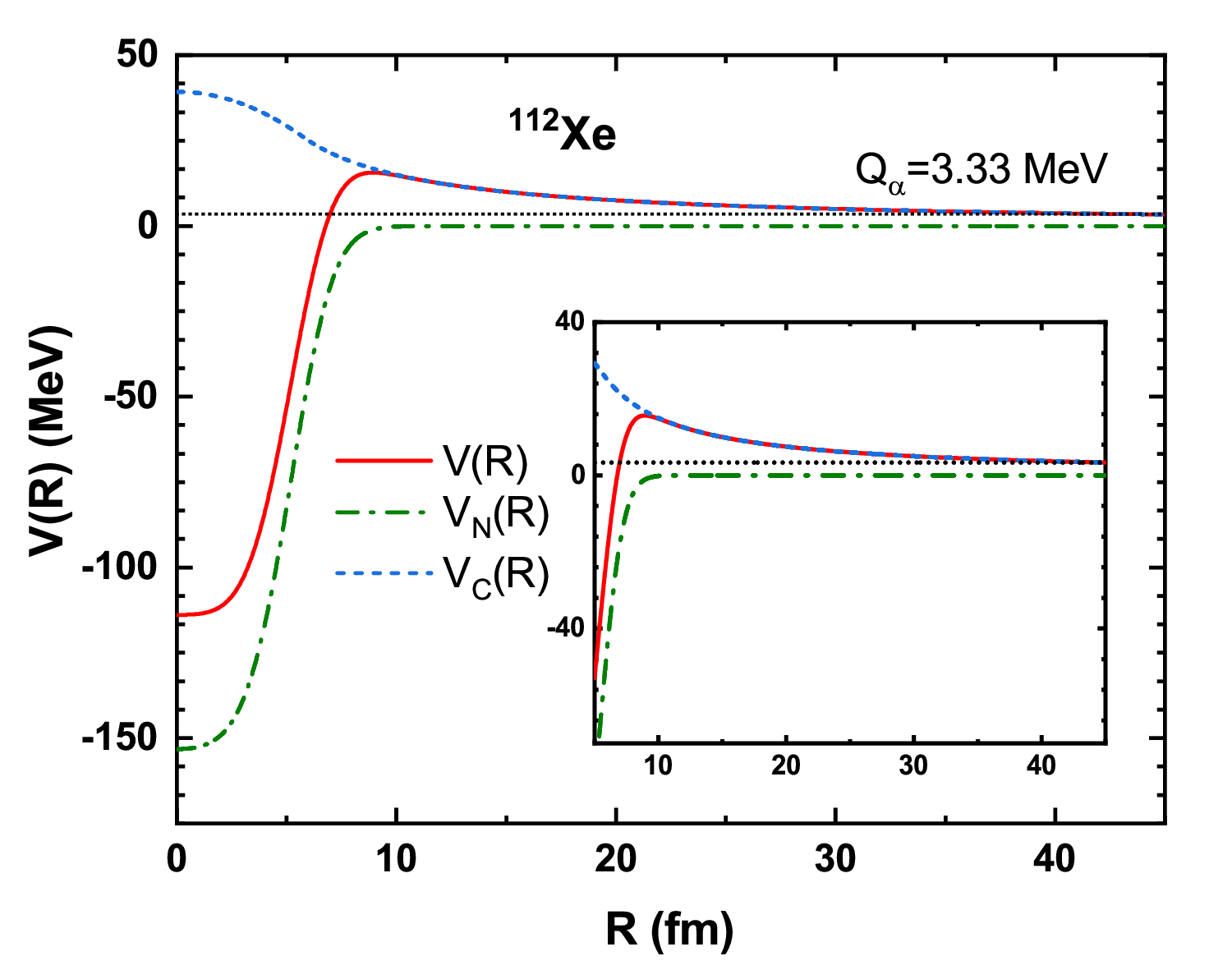}
	\end{center}
	\caption{(Color online) Total interaction potential $V(R)$ (red solid curve), nuclear potential $V_{N}\left( R \right )$ (green dot-dash curve) and Coulomb potential $V_{c}\left( R \right )$ (blue dash curve) for $^{112} $Xe, respectively.}
	\label{Fig.1}
\end{figure}

Next, taking $^{112}$Xe as an example, the total interaction potential $V(R)$, nuclear potential $V_{N}\left( R \right )$ and Coulomb potential $V_{c}\left( R \right )$ versus $R$ are obtained, which are plotted in Fig. \ref{Fig.1}. From Fig. \ref{Fig.1}, we can see that the well depth of $V(R)$ is about -115 MeV. Moreover, by the equation \textit{V}(\textit{R})=\textit{Q}$_{\alpha }^{\text{expt.}}$=3.33 MeV, the second and third turning points $R_{2} $ and $R_{3} $ are determined as 6.97 fm and 44.93 fm, respectively. The second turning point $R_{2} $ lying within the nuclear surface region indicates that the $\alpha$-particle is preformed inside the nucleus before penetrating the barrier through quantum tunneling effect. In $\alpha$-decay half-life calculation, the Bohr-Sommerfeld quantization condition in Eq. \eqref{Bohr} is used. The renormalization factor $\lambda $ is determined as 0.60. Therefore, the calculated $\alpha$-decay half-life is $ 4.27\times 10^{2} $ s, which is in good agreement with the corresponding experimental value $T_{1/2 }^{\mathrm{expt.} } = 2.25\times 10^{2}$ s.\\\vspace{-3ex}

\begin{table}[!h]
	\centering
	\caption{Comparison of experimental and calculated $\alpha$-decay half-lives of the nuclei approaching $N=Z$ (in seconds). The spin-parity $J^{\pi } $, the experimental $\alpha$-decay energies $Q_{\alpha }^{\text{expt.}} $ (in MeV) and the experimental half-lives are taken from Refs.~\cite{AME2021a,AME2021b}.}
	\label{tab2}
	
	\begin{threeparttable}[b]
		\begin{tabular}{c c c c c c c c c}
			\hline\hline
			Emitters&$J_{i}^{\pi }$& $J_{f}^{\pi }$&$l_{min}$ & $Q_{\alpha }^{\text{expt.}} $(MeV) & $P_{\alpha }  $&$T_{1/2 }^{\mathrm{expt.} } $(s)& $T_{1/2 }^{\mathrm{cal.} } $(s)& $\log_{10}{HF}   $   \\
			\hline
			$Z=52$&  &  &  &  &  &  &  & \\
			$^{104}\mathrm{Te} \to ^{100}\mathrm{Sn}+  \alpha $&$0^{+  } $&$0^{+  } $ &0& 5.1 &  $0.305^{*}$ & $<4\times 10^{-9} $ & $5.84\times 10^{-8} $& $<-1.16$  \\
			$^{105}\mathrm{Te} \to ^{101}\mathrm{Sn}+  \alpha $&$7/2^{+  } $&$7/2^{+  } $ &0& 5.069 &  $0.173^{*}$ & $6.33 \times10^{-7} $ & $1.22\times 10^{-7} $& 0.72   \\
			$^{106}\mathrm{Te} \to ^{102}\mathrm{Sn}+  \alpha $&$0^{+  } $&$0^{+  } $ &0& 4.29 & 0.248  & $7.8 \times10^{-5} $ & $1.12\times 10^{-4} $ & $-0.16$  \\
			$^{107}\mathrm{Te} \to ^{103}\mathrm{Sn}+  \alpha $&$5/2^{+  } $&$5/2^{+  } $ &0& 4.01 & 0.110  & $4.6 \times10^{-3} $ & $5.45\times 10^{-3} $ &$-0.07$   \\
			$^{108}\mathrm{Te} \to ^{104}\mathrm{Sn}+  \alpha $&$0^{+  } $&$0^{+  } $ &0& 3.42 & 0.255  & $4.286 \times10^{0} $ & $5.42\times 10^{0} $ &$-0.10$   \\
			$^{109}\mathrm{Te} \to ^{105}\mathrm{Sn}+  \alpha $&$5/2^{+  } $&$5/2^{+  } $ &0& 3.198 & 0.128  & $1.128 \times10^{2} $ & $3.42\times 10^{-2} $ &$-0.48$ \\
			$Z=53$&  &  &  &  &  &  &  & \\
			$^{108}\mathrm{I} \to ^{104}\mathrm{Sb}+  \alpha $ &$1^{+  } $ &$/$  &$/$  & 4.099 & 0.094  & $2.653 \times10^{-2} $ & $8.78\times 10^{-3}$ &0.48 \\
			$^{109}\mathrm{I} \to ^{105}\mathrm{Sb}+  \alpha $ &$1/2^{+  },~3/2^{+  } $ & $5/2^{+  } $  & 2 & 3.918 & 0.130  & $6.629 \times10^{-1} $ & $1.58\times 10^{0}$  &$-0.38$  \\
			$^{110}\mathrm{I} \to ^{106}\mathrm{Sb}+  \alpha $ &$1^{+  } $ & $2^{+  } $  &  2& 3.58 & 0.070 &  $3.906\times10^{0} $ & $2.05\times 10^{2}$ &$-1.72$ \\
			$^{111}\mathrm{I} \to ^{107}\mathrm{Sb}+  \alpha $ &$5/2^{+  } $ & $5/2^{+  } $  & 0 & 3.275 & 0.135  & $2.841\times10^{3} $ & $4.41\times 10^{2}$  &0.81\\
			$^{113}\mathrm{I} \to ^{109}\mathrm{Sb}+  \alpha $ &$5/2^{+  } $ & $5/2^{+  } $  &  0& 2.707 & 0.138  & $1.994\times10^{7} $ & $1.83\times 10^{7}$ &0.04\\
			$Z=54$ &  &  &  &  &  &  &  & \\
			$^{108}\mathrm{Xe} \to ^{104}\mathrm{Te}+  \alpha $ &$0^{+  } $ & $0^{+  } $  &  0& 4.57 &  $0.332^{*}$ & $7.2\times10^{-5} $ & $5.94\times 10^{-5}$ &0.08  \\
			$^{109}\mathrm{Xe} \to ^{105}\mathrm{Te}+  \alpha $ &$7/2^{+  } $ & $7/2^{+  } $  &  0& 4.217 &  $0.190^{*}$ & $1.3\times10^{-2} $ & $4.26\times 10^{-3}$ &0.48\\
			$^{110}\mathrm{Xe} \to ^{106}\mathrm{Te}+  \alpha $ &$0^{+  } $ & $0^{+  } $  &  0& 3.872 & 0.259  & $1.453\times10^{-1} $ & $1.97\times 10^{-1}$ &$-0.13$\\
			$^{111}\mathrm{Xe} \to ^{107}\mathrm{Te}+  \alpha $ &$5/2^{+  } $ & $5/2^{+  } $  &  0& 3.71 & 0.106  & $7.115\times10^{0} $ & $3.87\times 10^{0}$ &0.26\\
			$^{112}\mathrm{Xe} \to ^{108}\mathrm{Te}+  \alpha $ &$0^{+  } $ & $0^{+  } $  &  0& 3.33 & 0.276&  $2.25\times10^{2} $ & $4.27\times 10^{2}$ &$-0.28$ \\
			$Z=55$ &  &  &  &  &  &  &  &\\
			$^{112}\mathrm{Cs} \to ^{108}\mathrm{I}+  \alpha $ &$1^{+  } $ & $1^{+  } $  &  0& 3.93 &$0.086^{*}$& $>1.885\times10^{-1} $ & $1.12\times 10^{0}$ & $>-0.77$\\
			$^{114}\mathrm{Cs} \to ^{110}\mathrm{I}+  \alpha $ &$1^{+  } $ & $1^{+  } $  &  0& 3.36 &0.065&  $3.167\times10^{3} $ & $5.29\times 10^{3}$ &$-0.22$ \\
			$Z=56$&  &  &  &  &  &  &  & \\
			$^{114}\mathrm{Ba} \to ^{110}\mathrm{Xe}+  \alpha $ &$0^{+  } $ & $0^{+  } $  &  0& 3.592 &0.271& $5.111\times10^{1} $ & $1.52\times 10^{2}$ &$-0.47$ \\
			\hline\hline
		\end{tabular}
		\begin{tablenotes}
			\item[*] Calculated by using the WS4 mass model table~\cite{WS4}.
		\end{tablenotes}
	\end{threeparttable}
	
\end{table}

Encouraged by the good agreement between the calculated $\alpha$-decay half-life of $^{112}$Xe and the experimental data, we extend our model to calculate the $\alpha$-decay half-lives for more nuclei around $N=Z$, including the Te, I, Xe, Cs and Ba isotopes, which are listed in Table \ref{tab2}. The first column of Table \ref{tab2} corresponds to the transitions between initial and final states of $\alpha$ emitters. The second and third columns show the spin-parity of the parent and daughter nuclei, respectively. The fourth column represents the minimal orbital angular momenta $\left ( l_{min}  \right ) $ carried by the $\alpha$-particle, which are determined by the spin-parity selection rule. If the $J^{\pi } $ values of the parent or daughter nuclei are unknown, the $ l_{min}$ values are selected as 0. The experimental $Q_{\alpha }$ values~\cite{AME2021a} are listed in the fifth column. The preformation factors $P_{\alpha } $ are shown in column 6. The 7th and 8th columns are experimental and calculated $\alpha$-decay half-lives, respectively. The logarithm hindrance factors $\log_{10}{HF}   $ ($\log_{10}{HF}=\log_{10}{T_{1/2}^{\mathrm{expt.} } } - \log_{10}{T_{1/2}^{\mathrm{cal.} } }$) are listed in the last column.\\\vspace{-3ex}

\begin{figure}[!h]
	\begin{center}
		\includegraphics[width=12cm]{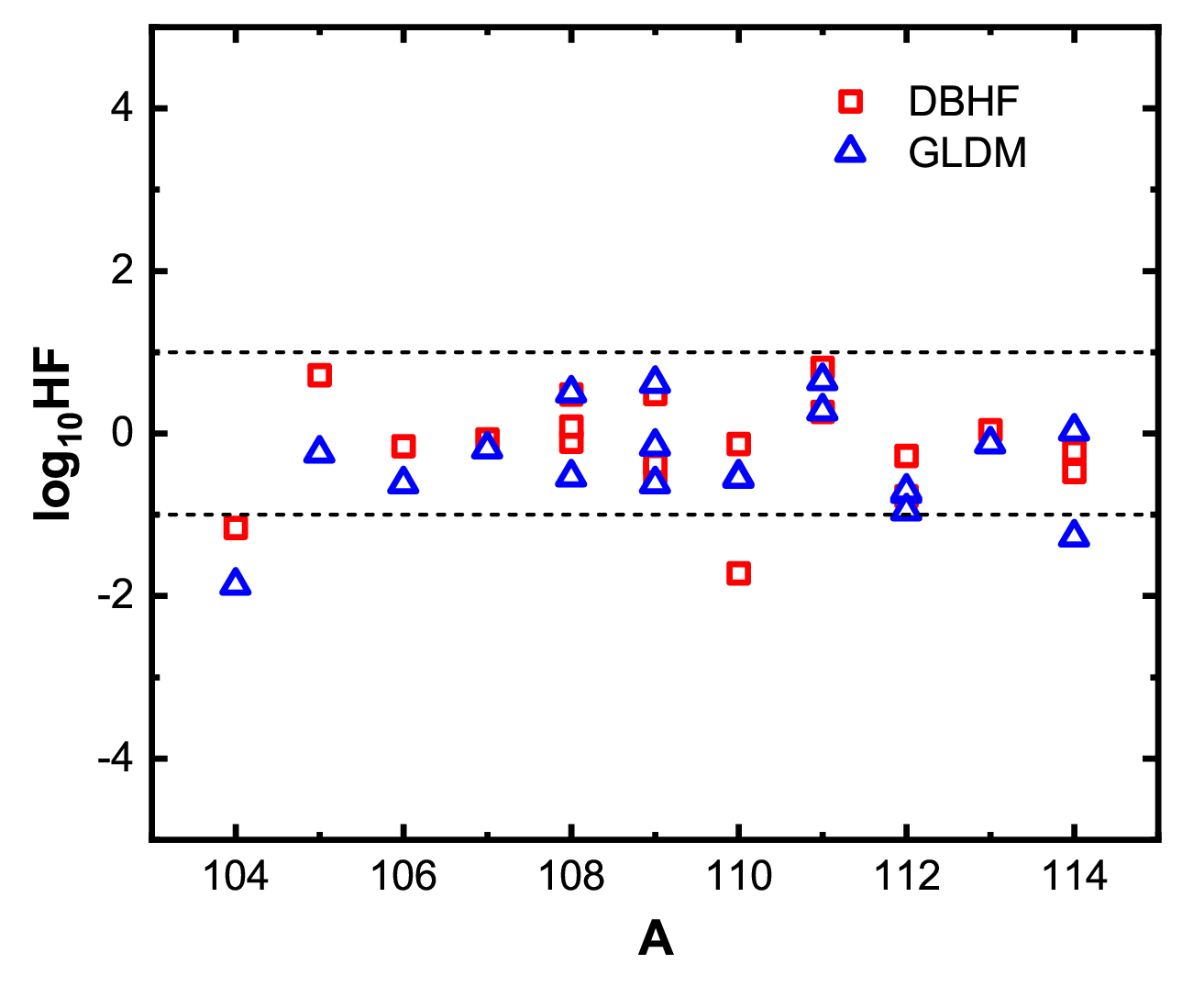}
	\end{center}
	\caption{(Color online) The logarithm hindrance factors $\log_{10}{HF} $ for the nuclei around $N=Z$. The red squares and blue triangles represent the results from the DBHF and the GLDM~\cite{Ghar22}, respectively.}
	\label{Fig.2}
\end{figure}

To show the agreement between the calculated $\alpha$-decay half-lives and experimental data more clearly, the $\log_{10}{HF}$ values are plotted in Fig. \ref{Fig.2}. Meanwhile, the $\log_{10}{HF}$ values from the generalized liquid drop model (GLDM)~\cite{Ghar22} are also shown. It is generally believed that if the $\log_{10}{HF}$ value is within a factor of 1.0, the calculated half-lives will be in agreement with the experimental data~\cite{Qi2024,Qi2025}. As can be seen from Fig. \ref{Fig.2}, more $\log_{10}{HF}$ values within the DBHF model fall between -1.0 and 1.0. This suggests that the DBHF calculations are in better agreement with the experimental data. Moreover, the success of the DBHF model indicates the rationality of the effective $n-n$ interaction obtained within the $G$ matrix starting from a bare $n-n$ interaction. To quantitatively compare the accuracies of the two models, the standard deviation $\sqrt{\bar{\sigma ^{2} } }$ between the calculated $\alpha$-decay half-lives and the experimental data is calculated by

\begin{equation}  \label{sigma2}
	\sqrt{\bar{\sigma ^{2} } } =\left [\frac{1}{n}\sum_{i=1}^{n}\left ( \log_{10}{T_{1/2,i}^{\mathrm{expt.} } } -\log_{10}{T_{1/2,i}^{\mathrm{cal.} } } \right )  ^{2}   \right  ] ^{1/2}.
\end{equation}

Within Eq. \eqref{sigma2}, the calculated $\sqrt{\bar{\sigma ^{2} } }$ values extracted by the DBHF and the GLDM are 0.62 and 0.78, respectively. This further suggests that the accuracy of the DBHF is higher than that of the GLDM. In addition, we find that the nuclear stability is enhanced with the increase of $N$ in each isotopic chain based on the experimental half-lives, which is attributed to the increase of the symmetry energy with $N$~\cite{Wang14}.\\\vspace{-3ex}

\begin{table}[h]
	\centering
	\caption{The predicted $\alpha$-decay half-lives of the nuclei approaching $N=Z$ (in seconds). The spin-parity $J^{\pi } $ and $Q_{\alpha } $ values (in MeV) are taken from Refs.~\cite{AME2021a,AME2021b}.}
	\label{tab3}
	
	\begin{threeparttable}[b]
		
		\begin{tabular}{c c c c c c c}
			\hline\hline
			Emitters&$J_{i}^{\pi }$& $J_{f}^{\pi }$&$l_{min}$& $Q_{\alpha } $(MeV)& $P_{\alpha } $& $T_{1/2 }^{\mathrm{cal.} } $(s)   \\
			\hline
			$^{110}\mathrm{Te} \to ^{106}\mathrm{Sn}+  \alpha $&$0^{+  } $&$0^{+  } $ &0& 2.699 & 0.239 & $2.31\times 10^{6} $  \\
			$^{111}\mathrm{Te} \to ^{107}\mathrm{Sn}+  \alpha $&$5/2^{+  } $&$5/2^{+  } $ &0& 2.500 & 0.119  & $4.04 \times10^{8} $    \\
			$^{112}\mathrm{Te} \to ^{108}\mathrm{Sn}+  \alpha $&$0^{+  } $&$0^{+  } $ &0& 2.078 & 0.226 & $2.67 \times10^{13} $    \\
			$^{113}\mathrm{Te} \to ^{109}\mathrm{Sn}+  \alpha $&$7/2^{+  } $&$5/2^{+  } $ &2& 1.858 & 0.096  & $3.36 \times10^{18} $     \\
			$^{114}\mathrm{Te} \to ^{110}\mathrm{Sn}+  \alpha $&$0^{+  } $&$0^{+  } $ &0& 1.527 & 0.223  & $1.15 \times10^{23} $     \\
			
			$^{106}\mathrm{I} \to ^{102}\mathrm{Sb}+  \alpha $ & &  &  & 5.380 &$ 0.138^{*}$ & $4.30 \times10^{-8} $  \\
			$^{107}\mathrm{I} \to ^{103}\mathrm{Sb}+  \alpha $ &$5/2^{+  }$ & $5/2^{+  } $  &0& 4.820&0.207 &  $2.68 \times10^{-6} $   \\
			$^{114}\mathrm{I} \to ^{110}\mathrm{Sb}+  \alpha $ &$1^{+  } $ & $3^{+  } $  &  2& 2.386 & 0.055 & $2.65\times10^{12} $   \\
			$^{115}\mathrm{I} \to ^{111}\mathrm{Sb}+  \alpha $ &$5/2^{+  } $ & $5/2^{+  } $  & 0 & 2.070 & 0.146 &  $4.18\times10^{14} $   \\
			
			$^{113}\mathrm{Xe} \to ^{109}\mathrm{Te}+  \alpha $ &$5/2^{+  } $ & $5/2^{+  } $  &  0& 3.087 & 0.137 &  $5.40\times10^{4} $   \\
			$^{114}\mathrm{Xe} \to ^{110}\mathrm{Te}+  \alpha $ &$0^{+  } $ & $0^{+  } $  &  0& 2.719 & 0.268 &  $4.43\times10^{7} $  \\
			$^{115}\mathrm{Xe} \to ^{111}\mathrm{Te}+  \alpha $ &$5/2^{+  } $ & $5/2^{+  } $  &  0& 2.506 & 0.132 & $1.30\times10^{10} $  \\
			$^{116}\mathrm{Xe} \to ^{112}\mathrm{Te}+  \alpha $ &$0^{+  } $ & $0^{+  } $  &  0& 2.096 & 0.256 &  $8.72\times10^{14} $  \\
			
			$^{111}\mathrm{Cs} \to ^{107}\mathrm{I}+  \alpha $ & $3/2^{+  } $ &  $5/2^{+  } $  & 2 & 4.110& $0.183^{*}$& $1.65\times10^{0} $ \\
			$^{113}\mathrm{Cs} \to ^{109}\mathrm{I}+  \alpha $ &$3/2^{+  } $ & $1/2^{+  },~3/2^{+  } $  &  0& 3.483 &0.134&  $3.60\times10^{2} $  \\
			$^{115}\mathrm{Cs} \to ^{111}\mathrm{I}+  \alpha $ & $9/2^{+  } $ &  $5/2^{+  } $  & 2 & 2.830&0.149&  $1.12\times10^{9} $ \\
			$^{116}\mathrm{Cs} \to ^{112}\mathrm{I}+  \alpha $ & $1^{+  } $ &  $1^{+  } $  & 0 & 2.600&0.068&  $1.62\times10^{10} $ \\
			$^{117}\mathrm{Cs} \to ^{113}\mathrm{I}+  \alpha $ & $9/2^{+  } $ &  $5/2^{+  } $  & 2 & 2.200&0.143&  $1.24\times10^{16} $ \\
			
			$^{113}\mathrm{Ba} \to ^{109}\mathrm{Xe}+  \alpha $ & $5/2^{+  } $& $7/2^{+  } $  & 2 & 4.040 &$ 0.195^{*}$& $1.24\times10^{1} $  \\
			$^{115}\mathrm{Ba} \to ^{111}\mathrm{Xe}+  \alpha $ & $5/2^{+  } $& $5/2^{+  } $  & 0 & 3.180 &0.133&  $2.74\times10^{5} $  \\
			$^{116}\mathrm{Ba} \to ^{112}\mathrm{Xe}+  \alpha $ & $0^{+  } $& $0^{+  } $  & 0 & 3.220&0.231& $7.56\times10^{4} $  \\
			$^{117}\mathrm{Ba} \to ^{113}\mathrm{Xe}+  \alpha $ & $3/2^{+  } $& $5/2^{+  } $  & 2 & 2.320 &0.160&  $2.45\times10^{15} $  \\
			$^{118}\mathrm{Ba} \to ^{114}\mathrm{Xe}+  \alpha $ & $0^{+  } $& $0^{+  } $  & 0 & 2.460 &0.221&  $1.24\times10^{12} $  \\
			
			\hline\hline
		\end{tabular}
		
	\end{threeparttable}

\end{table}

Driven by the success of the DBHF model, we attempt to predict the $\alpha$-decay half-lives for the nuclei around $N=Z$ that are unavailable experimentally, which are listed in Table \ref{tab3}. These predicted $\alpha$-decay half-lives are useful for identifying the new nuclides or isotopes in future experiments. \\\vspace{-3ex}

\begin{figure}[!h]
	\begin{center}
		\includegraphics[width=16cm]{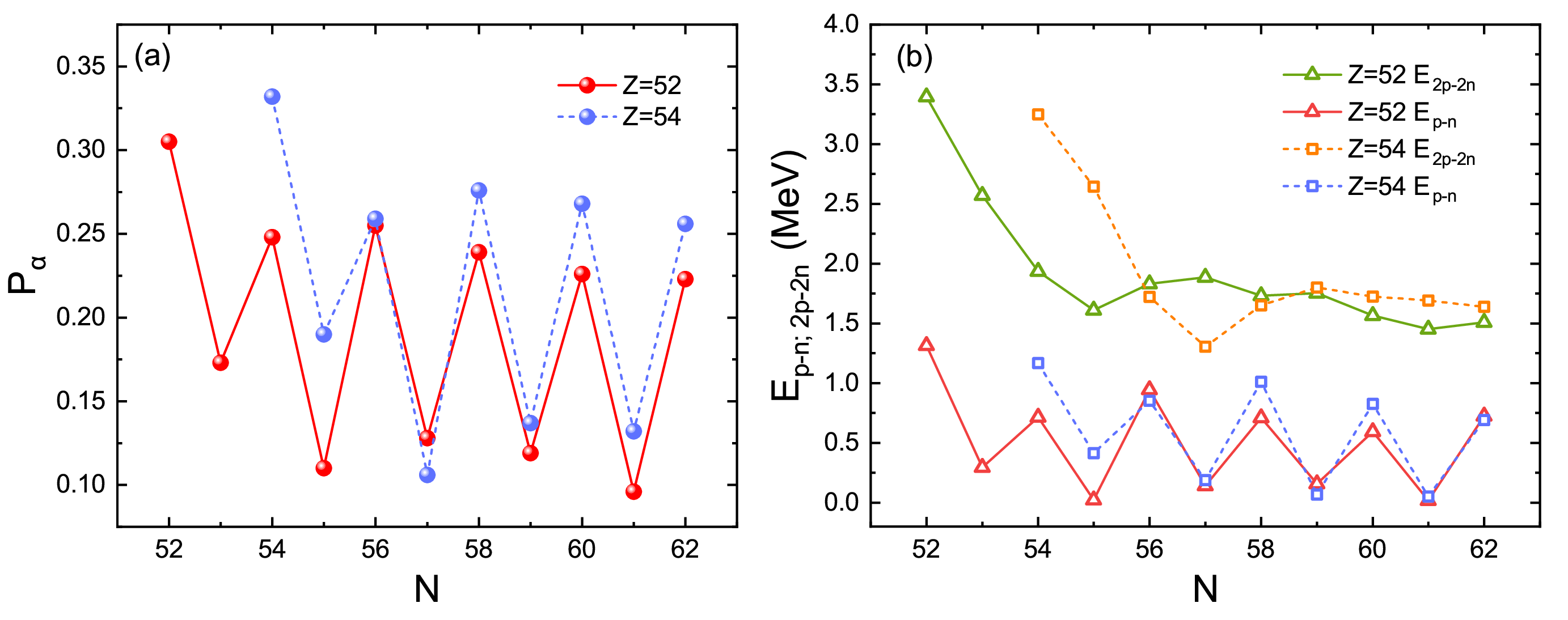}
	\end{center}
	\caption{(Color online) (a) The $P_{\alpha }$ values of $Z=52$ and $Z=54$ isotopes versus $N$. (b) Same as (a), but for the proton-neutron correlation energy $E_{p-n}$ and two protons-two neutrons correlation energy $E_{2p-2n}$, respectively.}
	\label{Fig.3}
\end{figure}

In the calculations on $\alpha$-decay half-lives, $P_{\alpha }$ extracted by the CFM is taken into account. From Tables I and II, it is seen that the order of magnitude of $P_{\alpha }$ of most nuclei is $10^{-2}$. It is larger than that of the nuclei near $^{208}$Pb, which is the same as some previous work~\cite{Wan2021,Wang14,Deng2022,Clark2020}. In our previous studies~\cite{Wang14,Wang14b}, we pointed out that $\alpha$-cluster preformation is not just dependent on a mean field including the pairing interaction. The high order correlations between nucleons plays an important role in $\alpha$-cluster preformation. The correlation between protons and neutrons is one of the most important high order correlations~\cite{Kane03,Deng2022,Ahmed2013,Kaneko99,Brenn90}. So, the correlation between protons and neutrons is usually applied to analyze the $\alpha$-cluster preformation. In Fig. \ref{Fig.3}(a), we show the $P_{\alpha }$ values of $Z=52$ and $Z=54$ isotopes as functions of $N$. It is seen that the $P_{\alpha }$ values increase gradually when the nuclei move towards the $N=Z$ line. This suggests that the $P_{\alpha }$ values are enhanced by the correlation between protons and neutrons occupying the similar single-particle states. This is a reason why the superallowed $\alpha$-decay near doubly magic nucleus $^{100}$Sn occurs. Moreover, the $P_{\alpha }$ values of odd-A nuclei are smaller than those of the neighboring even-even nuclei, which is usually named as the odd-even effect. To gain insight into the evolution of $P_{\alpha }$ with $N$, we calculate the proton-neutron correlation energy $E_{p-n}$ and two protons-two neutrons correlation energy $E_{2p-2n}$ by~\cite{Kane03,Deng2022,Ahmed2013}

\begin{equation}  \label{Ep-n}
	E_{p-n}=B\left ( A,~Z \right ) +B\left ( A-2,~Z-1 \right ) - B\left ( A-1,~Z-1 \right )-B\left ( A-1,~Z \right ),
\end{equation}

\begin{equation}  \label{E2p-2n}
	E_{2p-2n}=B\left ( A,~Z \right ) +B\left ( A-4,~Z-2 \right ) - B\left ( A-2,~Z-2 \right )-B\left ( A-2,~Z \right ).
\end{equation}

Here, the $B(Z, N)$ values are taken from the AME2020 atomic mass table~\cite{AME2021a} or the WS4 mass table~\cite{WS4}. The $E_{p-n}$ and $E_{2p-2n}$ values of $Z=52$ and $Z=54$ isotopes versus $N$ are plotted in Fig. \ref{Fig.3}(b). As can be seen from Fig. \ref{Fig.3}(b), the $E_{2p-2n}$ values are enhanced with the decrease of $N$, which is the same as the evolution of $P_{\alpha }$ with $N$ plotted in Fig. \ref{Fig.3}(a). Meanwhile, it is seen that the $E_{2p-2n}$ values are much larger than the $E_{p-n}$ values. It is suggested that for the $\alpha$-particle preformation, two protons-two neutrons interaction plays a more important role than the proton-neutron interaction, which is the same as the conclusion of a recent study~\cite{Deng2022}. Furthermore, it is observed that $E_{p-n}$ shows the odd-even staggering with $N$, which results in the odd-even effect of $P_{\alpha }$ shown in Fig. \ref{Fig.3}(a).


\section{Conclusion}\label{sec4}
In this work, an effective microscopic $n-n$ interaction through effective meson exchange was obtained within the DBHF $G$ matrix starting from a bare $n-n$ interaction is used to investigate the $\alpha$-decay half-lives of nuclei near the $N=Z$ line. For $\alpha$-decay half-life calculations, the $\alpha$-cluster preformation factor $P_{\alpha }$ extracted by the CFM framework is taken into account. We found that the calculated $\alpha$-decay half-lives agree with experimental data well and its accuracy is higher than that of the GLDM. This demonstrates the success and reasonableness of the effective $n-n$ interaction and CFM. Then, we perform predictions on $\alpha$-decay half-lives for the nuclei that the experimental data is not available. These predictions are useful for identifying the new nuclides or isotopes in future experiments. Next, the evolution of $P_{\alpha }$ with $N$ is studied by using the proton-neutron correlation energy $E_{p-n}$ and two protons-two neutrons correlation energy $E_{2p-2n}$. It is found that the two protons-two neutrons interaction is more essential than the proton-neutron interaction for the $\alpha$-cluster preformation. Moreover, the odd-even staggering of $E_{p-n}$ leads to the odd-even staggering of $P_{\alpha }$. At last, it is necessary to point out that the cluster radioactivity~\cite{pei2006,wyz2021,Royer2001,QLJ2023}, proton radioactivity~\cite{wyz2017,San2017,Delion2006,Dongmeng2023} and two-proton radioactivity~\cite{wyz2023,wyz2024,Zhou2024,Zhu2022,Pfutzner2023} are important decay modes for unstable nuclei. Therefore, it is interesting to extend the $n-n$ effective interaction within the DBHF approach to study these radioactivities, which are the work in progress.

\section*{Acknowledgments} \label{sec7}
This work was supported by the S\&T Program of Hebei (Grant No. 236Z4601G), the Hebei Natural Science Foundation (Grant No. A2025210016), the Scientific Research Foundation for the Introducing Returned Overseas Chinese Scholars of Hebei Province (Grant No. C20230360), the Science and Technology Project of Hebei Education Department (Grant No. QN2023240) and the Key Laboratory of High Precision Nuclear Spectroscopy, Institute of Modern Physics, Chinese Academy of Sciences (Grant No. IMPKFKT2021002).



\end{document}